# MP-PCA denoising for diffusion MRS data: promises and pitfalls

Jessie Mosso[1,2,3], Dunja Simicic[1,2,3], Cristina Cudalbu[1,2,\*], Ileana O. Jelescu[4,\*]

1 CIBM Center for Biomedical Imaging, Switzerland
2 Animal Imaging and Technology, EPFL, Lausanne, Switzerland
3 LIFMET, EPFL, Lausanne, Switzerland
4 Lausanne University Hospital CHUV, Service de radiodiagnostic et radiologie interventionnelle, Switzerland

\*: equal contribution

## Abstract

Diffusion-weighted (DW) magnetic resonance spectroscopy (MRS) suffers from a lower signal to noise ratio (SNR) compared to conventional MRS owing to the addition of diffusion attenuation. This technique can therefore strongly benefit from noise reduction strategies. In the present work, the Marchenko-Pastur principal component analysis (MP-PCA) denoising is tested on Monte Carlo simulations and on *in vivo* DW-MRS data acquired at 9.4T in the rat brain. We provide a descriptive study of the effects observed following different MP-PCA denoising strategies (denoising the entire matrix versus using a sliding window), in terms of apparent SNR, rank selection, noise correlation within and across b-values and quantification of metabolite concentrations and fitted diffusion coefficients. MP-PCA denoising yielded an increased apparent SNR, a more accurate $B_0$ drift correction between shots, and similar estimates of metabolite concentrations and diffusivities compared to the raw data. No spectral residuals on individual shots were observed but correlations in the noise level across shells were introduced, an effect which was mitigated using a sliding window, but which should be carefully considered.

# 1. Introduction

Magnetic resonance spectroscopy (MRS) is a powerful technique that provides unique information about brain metabolite concentrations *in vivo*. Combined with diffusion weighting (DW), information on metabolites' diffusivities which are expected to reflect properties of the tissue microstructure can be extracted[1–8]. These properties include cell geometry, characteristic sizes of compartments, cytosol viscosity and molecular crowding. Unlike water, metabolites are naturally compartmentalized and probe the intracellular space almost exclusively. Some metabolites are even considered to be largely specific to glial cells, such as glutamine (Gln) or myo-inositol (Ins), some to neurons, such as N-acetyl aspartate (NAA) or glutamate (Glu)[4,5,9–11], while others are found in all cell types, such as creatine in all its forms[12] (total creatine: tCr). This intrinsic compartment specificity makes DW-MRS an extremely powerful tool to probe brain microstructure, in combination or in contrast to water diffusion MRI.

However, MRS is an inherently low signal-to-noise (SNR) technique due to the much lower concentration of metabolites relative to water, resulting in the need for substantial spectral averaging. For DW-MRS, the averaging is even heavier to compensate for diffusion attenuation, and acquisition times become prohibitively long to parse multiple diffusion-weightings (b-values), directions or diffusion times. DW-MRS data is typically acquired in single-voxel fashion. When fine spatial localization is required to study small structures, low SNR cannot be compensated by large voxel volumes. In this case, post-processing methods aiming to minimize the noise variance and its impact on the quantification of MRS signals are needed.

Several denoising schemes have been proposed, but remarkably none of them has been fully adopted by the MRS community[13–25]. Some of these denoising techniques, typically based on singular value decomposition (SVD) or another sparse representation such as Fourier space or wavelets[23,24], have been implemented for spectroscopic imaging data (MRSI)[16–21], and mainly in clinical applications. These methods rely on linear predictability, partial separability of spatial-temporal modes, or both, of such data[17–20]. In addition, constraints on the spatial distribution of the signal with specific regularization, such as total generalized variation (TGV), has shown to further enhance the SNR in MRSI reconstruction[20]. TGV regularization aims to denoise by enforcing smooth spatial variations, however with known limitations in terms of detecting focal pathology[22].

The main challenge of sparse representations such as SVD resides in the determination of the appropriate thresholds that separate the noise from the signal. In MRS, this arbitrary threshold can lead to possible elimination of spectral features that are on the same order of magnitude as noise components. Other approaches based on smoothing using splines, sliding window or Gaussian window lead to a deterioration of spectral/temporal resolution as well as artefactual auto-correlation[24]. Finally, deep learning approaches have been very recently suggested[26,27], but likely require more investigation to become robust.

One solution to choosing a threshold in sparse domain has been proposed recently, with the initial aim to denoise diffusion MRI data[28]. It is based on the Marchenko-Pastur principal component analysis (MP-PCA) technique, which exploits the fact that noise eigenvalues follow the asymptotic universal Marchenko-Pastur distribution, a result of the random matrix theory for noisy covariance matrices. This method thus provides a data-driven (more specifically, noise-driven) approach to distinguish noise from the signal components in SVD, since the cut-off is obtained by iteratively fitting the MP distribution to the tail of eigenvalues, and has shown its superiority to TGV for instance[28]. In practice, MP-PCA is suitable for the denoising of data with a high level of redundancy and a constant noise level across them. In the case of a diffusion MRI dataset for example, this could correspond to images acquired with different diffusion-weightings and directions. Since its initial development for diffusion MRI, its applications have been extended to functional MRI[29,30], $T_2$ relaxometry[31], preclinical $^1$H-MRSI[21] and $^{31}$P-MRSI[32].

The aim of the present study was to implement and test the potential of MP-PCA for denoising $^1$H DW-MRS. The performance of MP-PCA was tested at 9.4T using Monte Carlo simulations and *in vivo* experiments in the rat brain.

## 2. Methods

The following terminology will be used throughout the manuscript. The SNR referred to as *temporal SNR* is defined as the magnitude (absolute value) of the first complex point of the FID over one standard deviation (SD) of noise, taken on the real part of the FID tail (time points 1500 to 2048)[33]. The SNR referred to as *spectral SNR* or *SNR* corresponds to the SNR of NAA singlet at 2.01 ppm, defined as the NAA peak height taken on the magnitude spectra to avoid phasing and linewidth issues, over one standard deviation of noise taken in a noise-only region of the real part of the spectra (from 8.2 to 10.9 ppm).

The term *apparent SNR* will be used to refer to the SNR after denoising. The term *shot* will be used to refer to every complex FID in each shell, i.e. of a line of matrix $Z$, according to a recent consensus on terminology in MRS[33]. The terms *shell* will be used to designate a set of 100 (simulations) or 128 (*in vivo*) shots for a given b-value.

### 2.1. Theory

Let $Z$ be an initial noisy matrix in the temporal domain, $Z \in \mathcal{M}_{n \times m}(\mathbb{C})$, where $n$ is the number of shots, and $m$ is the number of time points in the FID signal:

$$Z = \tilde{Z} + \varepsilon$$

where $\tilde{Z} \in \mathcal{M}_{n \times m}(\mathbb{C})$ is the signal information and $\varepsilon \in \mathcal{M}_{n \times m}(\mathbb{C})$ the Gaussian, uncorrelated noise. For this section, we will assume that $2n < m$ and $2n \gg 1$ (asymptotic condition of the MP law). The real and imaginary parts of $Z$ are concatenated on the first dimension ($n$), and the resulting matrix $Y \in \mathcal{M}_{2n \times m}(\mathbb{R})$ is centered, such that:

$$X = Y - \mathbf{1}_{2n}^T \bar{Y}$$

where $X \in \mathcal{M}_{2n \times m}(\mathbb{R})$, $\bar{Y} \in \mathcal{M}_{1 \times m}(\mathbb{R})$ is the column-wise mean of $Y$ and $\mathbf{1}_{2n}^T$ is a column vector of $2n$ ones. Matrix $X$ is then decomposed using the singular value decomposition:

$$X = USV^T$$

where $U \in \mathcal{M}_{2n \times 2n}(\mathbb{R})$, $S \in \mathcal{M}_{2n \times m}(\mathbb{R})$ and $V \in \mathcal{M}_{m \times m}(\mathbb{R})$. Columns of $U$ are singular vectors of the first dimension (shots), columns of $V$ are singular vectors of the second dimension (time points) and $S$ contains the singular values of $X$, arranged in descending order, which are also the square root of the eigenvalues of $X^T X$. Since $X = Y - \mathbf{1}_{2n}^T \bar{Y}$, $\frac{1}{2n} X^T X$ is the covariance matrix of $Y$. The Marchenko-Pastur distribution is then fitted to the smallest non-zeros eigenvalues $\lambda$ of $\frac{1}{2n} X^T X$:

$$p(\lambda | \sigma, (2n-P)/m) = \begin{cases} \frac{\sqrt{(\lambda_+ - \lambda)(\lambda - \lambda_-)}}{2\pi \lambda \sigma^2 (2n-P)/m} & \text{if } \lambda_- \leq \lambda \leq \lambda_+ \\ 0 & \text{otherwise} \end{cases}$$

where $\sigma$ is the noise level estimated from the input matrix $X$, $P$ is the number of signal-carrying eigenvalues, $\lambda_-$ the smallest noise-related eigenvalue and $\lambda_+$ the largest. $P$ corresponds to the number of values $\lambda$ such that $\lambda \geq \lambda_+$, with $\lambda_+ = \sigma^2 \left(1 + \sqrt{\frac{2n-P}{m}}\right)^2$. The matrix $Y$ can then be approximated by:

$$\hat{Y} = U S_P V^T + \mathbf{1}_{2n}^T \bar{Y}$$

where $S$ has been truncated at rank $P$.

## 2.2. Monte Carlo simulations

Synthetic ¹H MR spectra were created (Matlab, MathWorks, Natick, MA, USA) to mimic experimental conditions in the rat brain (see Section 2.3 below). 19 metabolites, listed with their corresponding concentrations in **Table 1**, were simulated using NMRSCOPE-B from jMRUI[34], with published J-coupling and chemical shifts constants[35,36] and the SPECIAL sequence (9.4T, echo time (TE)=2.8ms). The lineshapes of the individual signals were constructed using a sum of 0.2 Hz Lorentzian and 1.8 Hz Gaussian apodizations, and a full macromolecule spectrum acquired *in vivo* (MM, 1.3mM) was included[37].

The free induction decays (FID) were generated with 2048 points. Diffusion weighting was simulated using Callaghan's model of diffusion in randomly oriented sticks[38], with ten b-values: 0.4, 1.5, 3.4, 6, 7.6, 13.4, 15.7, 20.8, 25.2, 33.3 ms/µm$^2$. Intra-stick free diffusion coefficients ranging from 0.265 to 0.67 µm$^2$/ms (**Table 1**) were attributed to the 19 metabolites and 0.005 µm$^2$/ms to the MM. Metabolite free diffusivities were set to be five times the apparent diffusion coefficient (ADC) of the ensemble of randomly-oriented cellular processes in the rodent brain from literature[5]. These values were retrospectively found to be in the same range as the intra-stick free diffusion coefficients estimated *in vivo* in the present work. A residual water signal was added to each spectrum (16 Hz Lorentzian line width, mono-exponential decay with apparent diffusivity 0.2 µm$^2$/ms, random phase). An additional 5 Hz Lorentzian line broadening was finally applied to all spectra. To simulate the full dataset for MP-PCA denoising (matrix $Z$), Gaussian noise was added to the real and imaginary parts of the FID, with a single shot temporal SNR of 13. One hundred noisy FIDs were generated for each b-value (constituting a "shell") and $B_0$ drifts (random -15/+15Hz drift) and phase distortions (random 0/30° phase) were added on individual shots, mimicking high SNR experimental *in vivo* DW-MRS data. The initial matrix $Z$ thus consisted of 1000 rows (10 b-values, 100 shots per b-value) and 2048 columns (FID time points).

Finally, the matrix $Z$ was generated 100 times with different noise realizations, water residual signal, $B_0$ drifts and phase distortions, and the effect of denoising on simulations was assessed in terms of variations across the MC iterations.

| Metabolite | Concentration (mM) | $D_{intra}$ (µm$^2$/ms) | Metabolite | Concentration (mM) | $D_{intra}$ (µm$^2$/ms) |
|---|---|---|---|---|---|
| Alanine (Ala) | 0,8 | 0,2695 | Lactate (Lac) | 0,8 | 0,65 |
| Ascorbate (Asc) | 1,5 | 0,3115 | N-acetylaspartate (NAA) | 9 | 0,4 |
| Aspartate (Asp) | 2 | 0,67 | scyllo-Inositol (Scyllo) | 0,1 | 0,3805 |
| Creatine (Cr) | 4 | 0,5 | Taurine (Tau) | 4,5 | 0,55 |
| Phosphocreatine (PCr) | 4,5 | 0,5 | Glucose (Glc) | 1,7 | 0,57 |
| gamma-Aminobutyric acid (GABA) | 1,6 | 0,378 | N-acetylaspartylglutamate (NAAG) | 0,3 | 0,4 |
| Glutamine (Gln) | 3 | 0,384 | Phosphatidylethanolamine (PE) | 0,5 | 0,318 |
| Glutamate (Glu) | 10 | 0,5 | Glycerophosphocholine (GPC) | 0,8 | 0,45 |
| Glutathione (GSH) | 1,5 | 0,2655 | Phosphocholine (PCho) | 0,2 | 0,45 |
| myo-inositol (mIns) | 6,5 | 0,45 | | | |

**Table 1** - Simulated metabolites with their respective concentrations and diffusion coefficients used in the MC simulations.

## 2.3. *In vivo* experiments

All experiments were approved by The Committee on Animal Experimentation for the Canton de Vaud, Switzerland. $^1$H DW-MRS acquisitions were performed on a horizontal actively shielded 9.4 Tesla system (Magnex Scientific, Oxford, UK) interfaced to a Varian Direct Drive console (Palo Alto, CA, USA) and equipped with 400mT/m gradients using a home-built 14 mm diameter surface $^1$H-quadrature transceiver.

Four adult male Wistar rats were scanned under isoflurane anesthesia (~1.5%). During the DW-MRS experiments, animals were placed in an in-house-built cradle, and their head was fixed in a stereotaxic system (bite bar and a pair of ear bars). The respiration rate and body temperature were monitored using a small-animal monitor system (SA Instruments, New York, NY, USA). Body temperature was measured with a rectal thermosensor and maintained at 37.7 ± 0.2 °C by warm water circulation.

First- and second-order shims were adjusted using FASTMAP[39], achieving water linewidths of 18-21 Hz in the volume of interest (VOI). DW-MRS data were acquired using a localized STEAM-based spectroscopic pulse sequence (TE/mixing time (TM)/repetition time (TR)=15/112/4000 ms) in a VOI of 162 to 245μl depending on the animal. The water signal was suppressed by using the VAPOR module interleaved with outer volume suppression blocks[40]. Diffusion gradients were applied simultaneously along three orthogonal directions ($\delta$=6 ms, $\Delta$=120 ms). A total of eleven $b$-values with 128 shots were acquired: 0.4, 1.5, 3.4, 6.0, 7.6, 9.3, 13.4, 15.7, 20.8, 25.2 and 33.3 ms/μm$^2$.

## 2.4. MP-PCA denoising

The raw data individual spectra in matrix $Z$ were first phase-corrected (maximization of the area of the metabolites region for each spectrum). The resulting complex-valued FID were split into real and imaginary parts and organized into a matrix $Y$ where the second dimension contained the time domain sampling and the first dimension a concatenation of all shots/b-values/real and imaginary parts. This was done in order to balance the number of rows with the number of columns and to increase the smallest dimension of $X$. The matrix $Y$ was centered column-wise and assigned to matrix $X$. A summary of the denoising strategies and of the study design is presented in **Figure 1**.

MP-PCA denoising performances were first tested on shots with no diffusion weighting ("single-shell") and different noise generations on the MC simulations, and compared to summation of the individual shots. A matrix $X$ of size 200x2048 was made of 100 single shots of the same shell (here: b=0), for multiple noise levels (SNR 13 – **Figure 2** and SNR 1, 2 and 5 – **Figure S 1**), as well as without or with phase/frequency drifts in the original spectrum (**Figure 2** and **Figure S 1**, respectively).

For denoising heterogeneous matrices $X$ composed of all b-values ("multi-shells"), three strategies were compared, both on MC simulations and on *in vivo* data:

1) **Multi-shell full matrix denoising - strategy 1**: MP-PCA denoising was performed on the full matrix. For MC simulations, the matrix $X$ to denoise was of size 2000x2048: 10 shells with 100 shots (i.e. noise realisations), and recreated 100 times. For *in vivo* data, the matrix $X$ to denoise was of size 2816x2048: 11 shells with 128 shots.

2) **Multi-shell sliding window (sw) denoising - strategy 2**: MP-PCA denoising was performed on a subset of the full matrix, using a sliding window of sub-blocks of three shells among all shells, and the denoised spectra output is selected when the shell is the middle of the 3-shell sub-block (similarly to dMRI procedure[28]). The first and last shells were selected together with the second shell and one before last from the first and last sub-blocks, respectively (**Figure 1**).

3) **Multi-shell sliding window denoising with half shots - strategy 3**: the same denoising procedure as strategy 2 was applied, but using only half of available shots (50 per b-value for MC simulations and 64 per b-value for *in vivo* acquisitions).

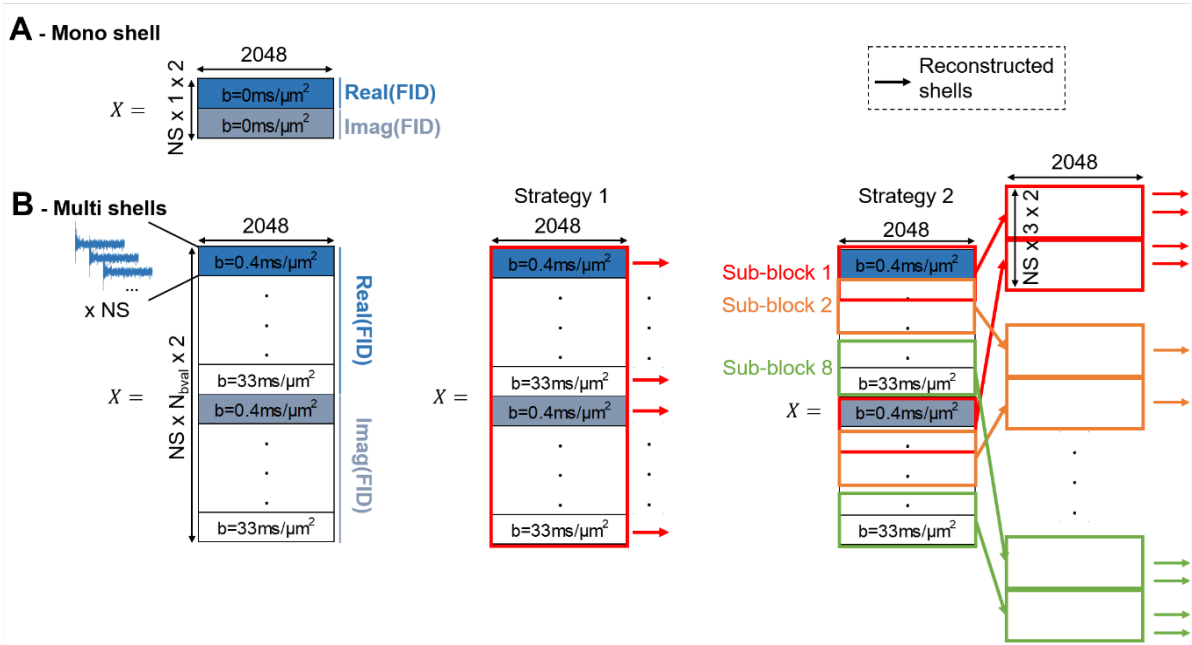

**Figure 1.** Study design and denoising strategies. **A**: Matrix organization for denoising a single-shell. This approach led to a similar result as summation of the shots on MC simulations (see **Figure 2**). **B**: Matrix organization for multi-shell full matrix (strategy 1) and sliding window denoising (strategy 2), the latter showing a reduced noise heterogeneity across shells. Strategy 3 (identical to strategy 2 with half the number of shots) is not displayed, showing similar results as strategy 3, yet with an increased number of outliers in the diffusion decay estimates. NS: number of shots, $N_{bval}$: number of b-values.

## 2.5. Quantification and modelling

Raw and denoised spectra were further corrected for $B_0$ drifts (alignment of the tCr peak at 3.03 ppm or NAA at 2.01 ppm in each spectrum to its position in the first spectrum after 8 Hz apodization) and summed (for each *b*-value). Metabolite concentrations were quantified using LCModel. The metabolites basis set was composed of the noiseless simulated signals for the MC study, and of spectra simulated using the acquisition parameters for the *in vivo* acquisitions, both basis sets containing an *in vivo*-acquired macromolecule signal. In addition, for the *in vivo* data, separately simulated MM and lipid components from LCModel were included to compensate for possible lipid contamination due to the large size of the voxel and its position close to the scalp[41,42]. The LCModel parameter controlling the baseline stiffness, DKNTMN, was set to 0.25 except for the quantification of the noiseless data in the simulations for which it was set to 1.

The randomly oriented stick model was fitted to the decay of each metabolite concentration as a function of b-value using a non-linear least squares algorithm in Matlab (*fit* function, *Trust-Region* method). The concentration decays as a function of b-value were fitted for each of the 100 MC iterations for simulations, and for each rat individually for the *in vivo* data. The mean estimated diffusion coefficients $D_{intra}$ with SD (across the 100 MC iterations or across the animals) were extracted. Percentage bias is reported for the concentrations and $D_{intra}$ (($Value_{method}$-$Value_{noiseless}$)/$Value_{noiseless}$).

Statistical tests were performed in RStudio (RStudio, PBC, Boston, MA). For simulations, $D_{intra}$ estimates based on raw and denoised data (from each denoising strategy) were compared to the $D_{intra}$ estimate from the noiseless data using a one-way ANOVA, and p-values were corrected for multiple comparisons with Dunnett's post-hoc test. For *in vivo* data, $D_{intra}$ estimates based on raw and denoised data (from each denoising strategy) were compared using a one-way ANOVA, and pairwise p-values were corrected for multiple comparisons with Tukey's post-hoc test. The following statistical significance values were used: * $p<0.05$, ** $p<0.01$, *** $p<0.001$, **** $p<0.0001$.

## 3. Results

The performance of the denoising strategies was assessed in terms of apparent SNR, spectral residuals (denoised summed spectra minus raw summed spectra for a given shell), rank selection, noise correlation within and across shells, as well as precision and accuracy of metabolite quantification for each b-value and of resulting diffusivity estimation.

### 3.1 Monte Carlo simulations

This section aims to study the effect of MP-CA denoising on simulated DW-MRS data while having access to the ground truth.

#### 3.1.1 Single-shell: MP-PCA denoising versus summation

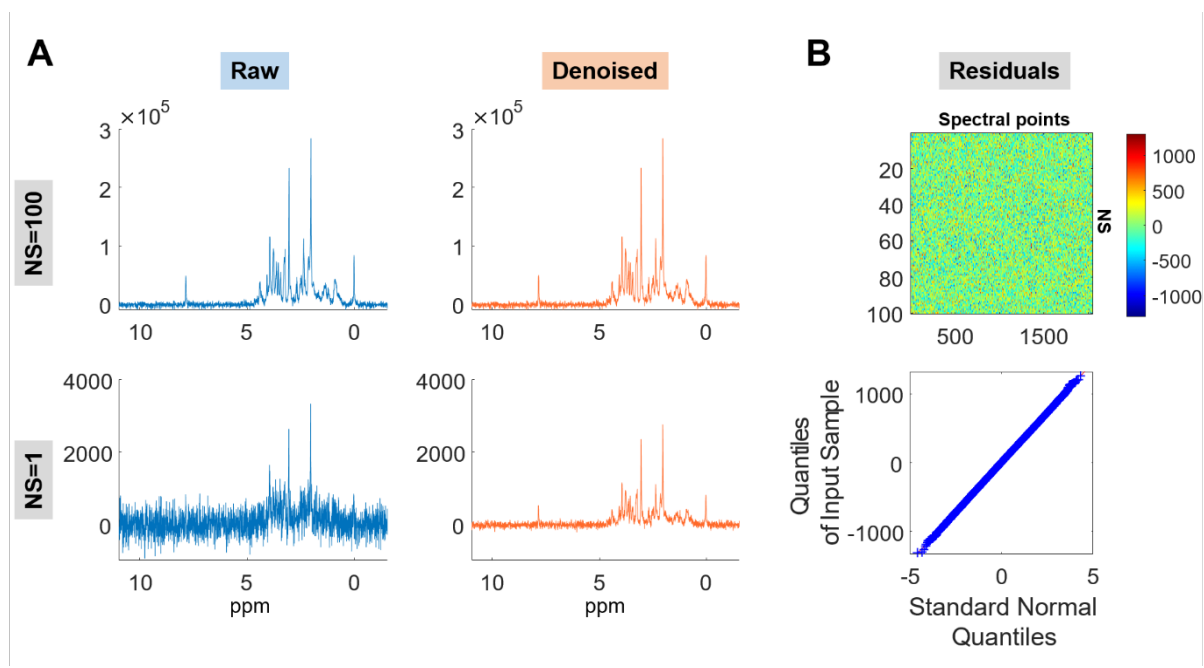

**Figure 2.** MP-PCA denoising performance on NS=100 shots of the same shell (with different noise realizations and no phase/frequency drifts). **A:** Raw (blue) and denoised (orange) spectra, of the summed 100 shots (top) and of a single shot (bottom): $SNR_{raw,100}$=101.9, $SNR_{raw,1}$=11.0, $appSNR_{dn,100}$=101.8, $appSNR_{dn,1}$=51.2. **B top:** Residuals (Denoised minus raw matrix) for the real part of the spectra. **B bottom:** Quantile-quantile (Q-Q) plot of the spectral residuals. Denoising a single-shell performs similarly to the summation of single shots (rank $P = 1$ selected by the MP fit) and yields to a Gaussian noise distribution after denoising. NS: number of shots.

**Figure 2** shows the performance of MP-PCA denoising on a single-shell matrix, i.e. N=100 shots of a spectrum with no diffusion weighting. Since summation (accumulation of spectra with different noise realisations but the same signal content) is a very efficient denoising strategy, it will be compared to MP-PCA. For a single-shell, denoising performs similarly to averaging on the summed spectra (NS=100, **Figure 2**A, top) and a rank $P = 1$ is selected by the MP fit. Single shots are also strongly denoised (NS=1, **Figure 2**A, bottom) but this representation should be handled with care since single shots are reconstructed from the entire denoised matrix and thus are not an equivalent representation of single shot raw data. The spectral residuals (100 shots x 2048 real spectral points) follow a Gaussian distribution and no structure in the metabolites' region was observed (**Figure 2**B). When phase and frequency drifts are applied across shots on the simulated spectrum, and at sufficiently high SNR, a rank $P > 1$ is retained by the MP fit (Supplementary **Figure S 1**B).

### 3.1.2 Multi-shell - strategy 1: MP-PCA denoising on the entire diffusion-weighted matrix

**Figure 3**A shows simulated diffusion-weighted spectra at 10 b-values. 11 principal components were retained by the MP fit (**Figure 3**B). The raw and denoised spectra for the two extreme b-values are shown in **Figure 3**C, for a single shot (bottom) and for the sum of the 100 shots (top). Denoising yields an improved spectral apparent SNR, on individual shots and on their sum. The central panel of **Figure 4**A shows that the noise level is non-uniform across shells after denoising with strategy 1, the shell

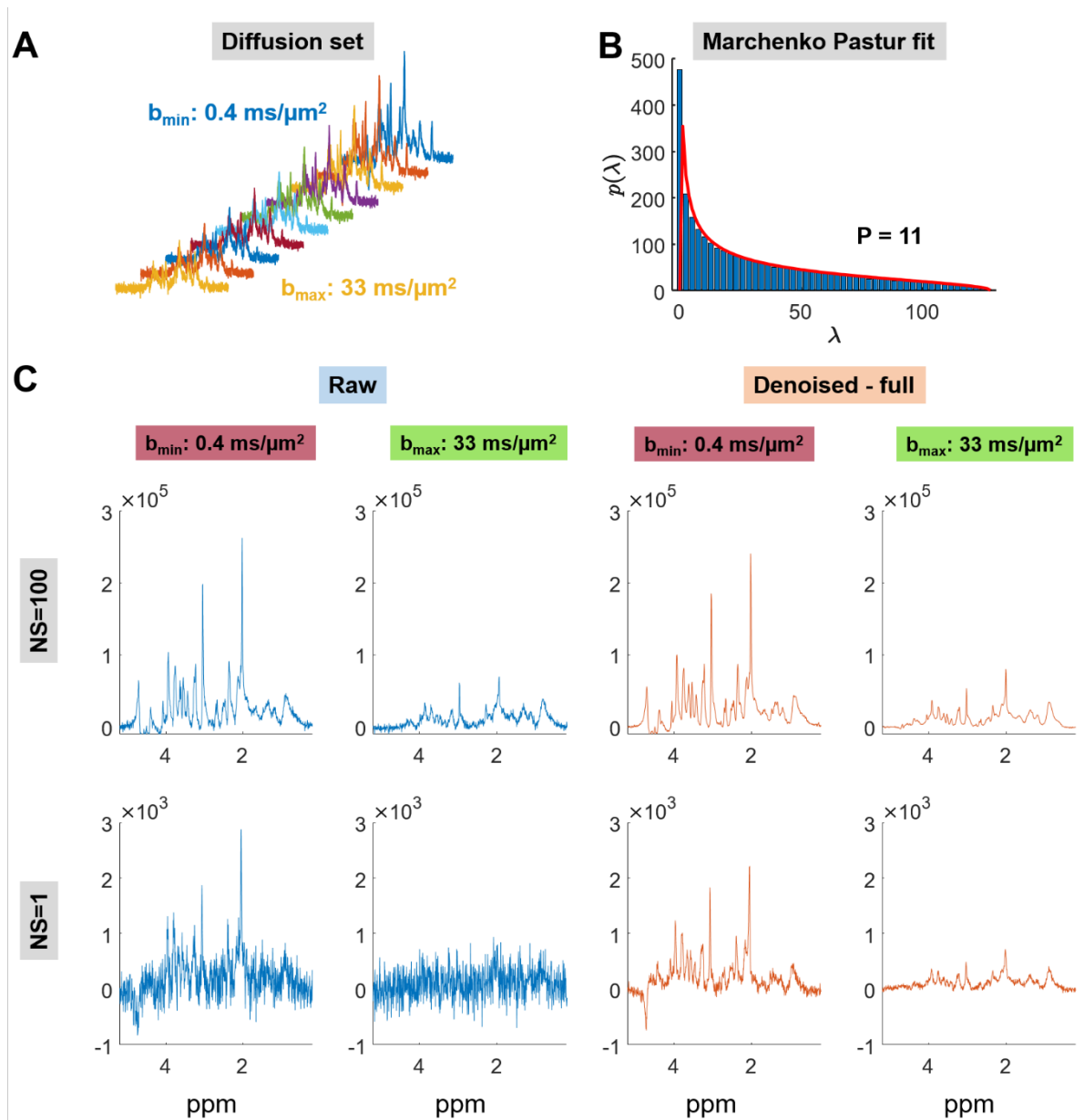

**Figure 3.** MP-PCA denoising performances on the full diffusion-weighted matrix $X$ made up of 10 shells, with 100 shots (NS) each (10x100x2 x 2048 FID points – 'x2' is for the concatenation of real and imaginary parts of the FID) – strategy 1. **A:** Simulated diffusion-weighted spectra at each b-value. **B:** Example MP fit on matrix $X$ for strategy 1 for one MC iteration. **C:** Example raw and denoised spectra, at low and high b-value, of the sum of the 100 shots (top) and of a single shot (bottom). $SNR_{raw,100,bmin}=90.4$, $SNR_{raw,1,bmin}=11.9$, $SNR_{raw,100,bmax}=24.0$, $SNR_{raw,1,bmax}=4.3$, $appSNR_{dn,100,bmin}=229.3$, $appSNR_{dn,1,bmin}=41.8$, $appSNR_{dn,100,bmax}=155.4$, $appSNR_{dn,1,bmax}=27.7$. Denoising improves apparent spectral SNR.

containing the higher b-value experiencing a stronger denoising effect, as evidenced by the ratio of spectral noise variances at $b_{min}$ and $b_{max}$.

### 3.1.3 Multi-shell - strategy 2 versus strategy 1

An alternative strategy of denoising using a sliding window of 3 shells, denoted as a "sub-block", is proposed (strategy 2), and aims at circumventing the effect of non-uniform noise level across shells, and compared to strategy 1. This resulted in minimal SNR heterogeneity within each sub-block on which the denoising was applied and is similar to what is used in dMRI[26] where the columns of matrix *X* are composed of a sliding spatial kernel of voxels. However, here we strive to reduce heterogeneity in the diffusion dimension (row-wise).

Although strategy 2 shows smaller noise reductions versus raw compared to strategy 1 (at $b_{min}$, 2.3 apparent SNR increase for strategy 2 versus 2.7 for strategy 1, at $b_{max}$, 3.6 apparent SNR increase for strategy 2 versus 6.8 for strategy 1, **Figure 4**C), it reduced the non-uniform noise levels across shells (**Figure 4**A). On the summed spectra: $\frac{\sigma_{bmin}}{\sigma_{bmax}} = 2.24$ for strategy 1 and $\frac{\sigma_{bmin}}{\sigma_{bmax}} = 1.65$ for strategy 2, whereas this ratio before denoising was close to 1 since single shots were created with the same noise level in each shell. The excessive noise reduction at high b-values (and potential wiping of signal) is also manifest, yet reduced with strategy 2. Noise levels on single shots display the same overall pattern as on the sum (**Figure 4**A, bottom). These observations suggest that some correlation is introduced in the noise, also shown in **Figure 5**.

The decreasing number of signal-carrying components retained by the MP fit as a function of sub-block number *(***Figure 4**B) highlights that, at low SNR (i.e. the noisiest sub-matrix, containing the highest b-values), less meaningful information can be separated from the noise (also shown in **Figure S 1**). In strategy 2, the apparent spectral SNR (**Figure 4**C) increases by a factor of 2.3 at $b_{min}$ and 3.7 at $b_{max}$ and follows a similar trend as in the raw data. However, it reaches a maximum for central b-values in strategy 1, possibly resulting from a "decay" of the apparent noise levels, as detailed in **Figure S 2**. The term SNR after denoising should be used carefully in the light of the noise correlations described below (**Figure 5**).

The summed residuals across shots (**Figure 4**D) show hardly any structure in the metabolites' region. The weak residuals around the NAA and Cr peaks may be caused by differences between the phase and frequency drift correction factors estimated from the raw or denoised data, or a change in linewidth after denoising, leading to spectral misalignment before summation[43].

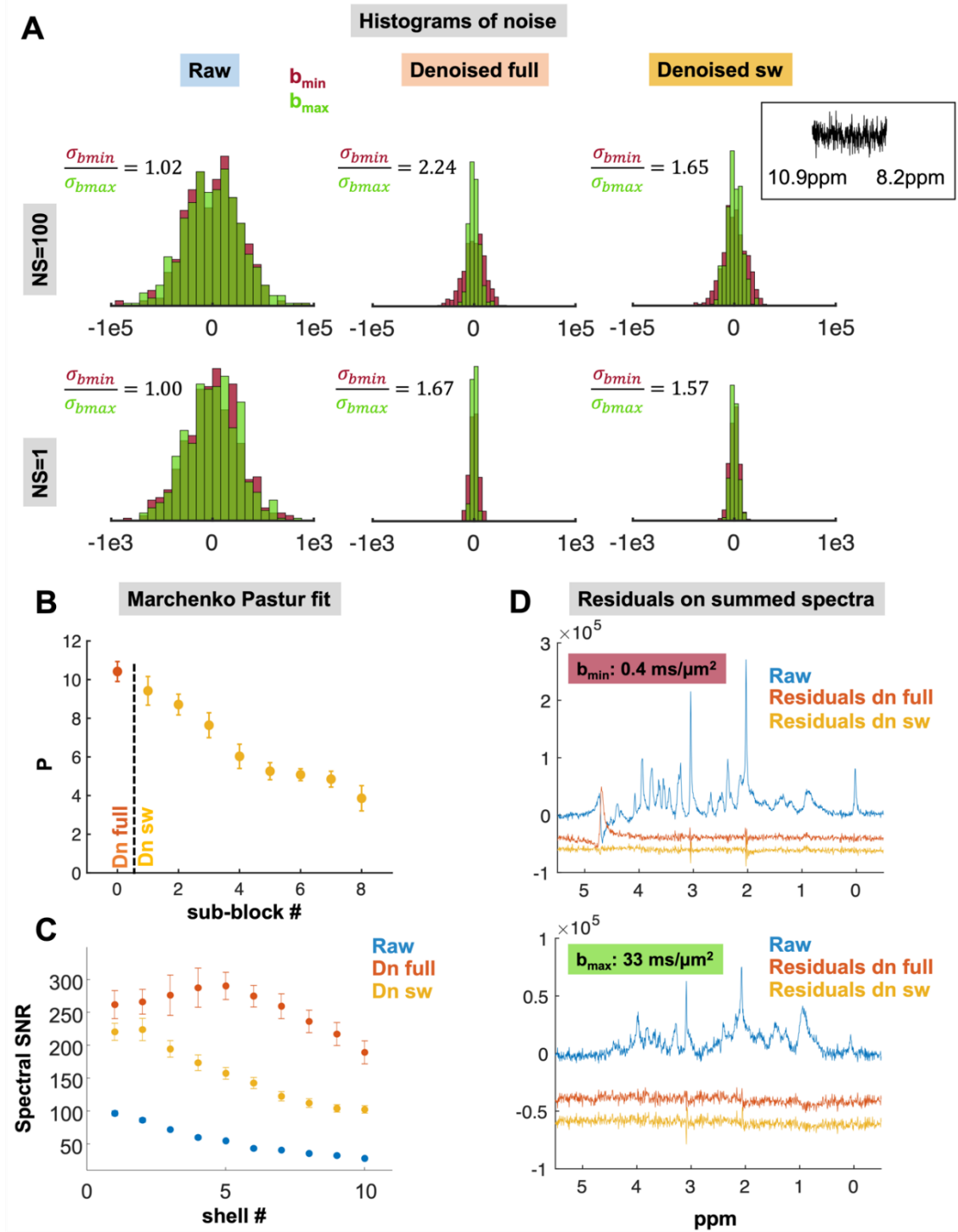

**Figure 4.** Comparison of MP-PCA denoising performance on the full matrix (strategy 1) or using a sliding window of 3 shells over the diffusion-weighted matrices (strategy 2). **A:** Histograms of spectral noise between 8.2 and 10.9 ppm (a region with no signals), for a single shot (bottom) and for the sum of the 100 shots (top), before and after denoising using strategies 1 and 2, for the lowest (red) and highest (green) b-values. The mean ratio across MC iterations of the noise level at $b_{min}$ over the one at $b_{max}$ is displayed in each case. Standard deviations across MC associated to the mean ratios displayed: for $NS = 100$, 0.04 (raw), 0.24 (dn full), 0.15 (dn sw), and for $NS = 1$, 0.05 (raw), 0.33 (dn full), 0.40 (dn sw). **B:** Number of principal components retained as signals (i.e. the rank P) by the MP fit, in strategy 1 (orange) and for each sub-block in strategy 2 (yellow), as mean and SD across MC iterations. **C:** Spectral (apparent) SNR on the summed spectra for each shell of raw and denoised data (strategy 1 & 2), as mean and SD across MC iterations. **D:** Spectral residuals on the summed spectra for the two denoising strategies at low (top) and high (bottom) b-values, shifted downwards for display. Both denoising strategies gave heterogeneous noise levels and increases in apparent SNR with no structure in spectral residuals. Strategy 2 mitigates some effects of strategy 1, namely the non-uniform SNR gain and variance across shells.

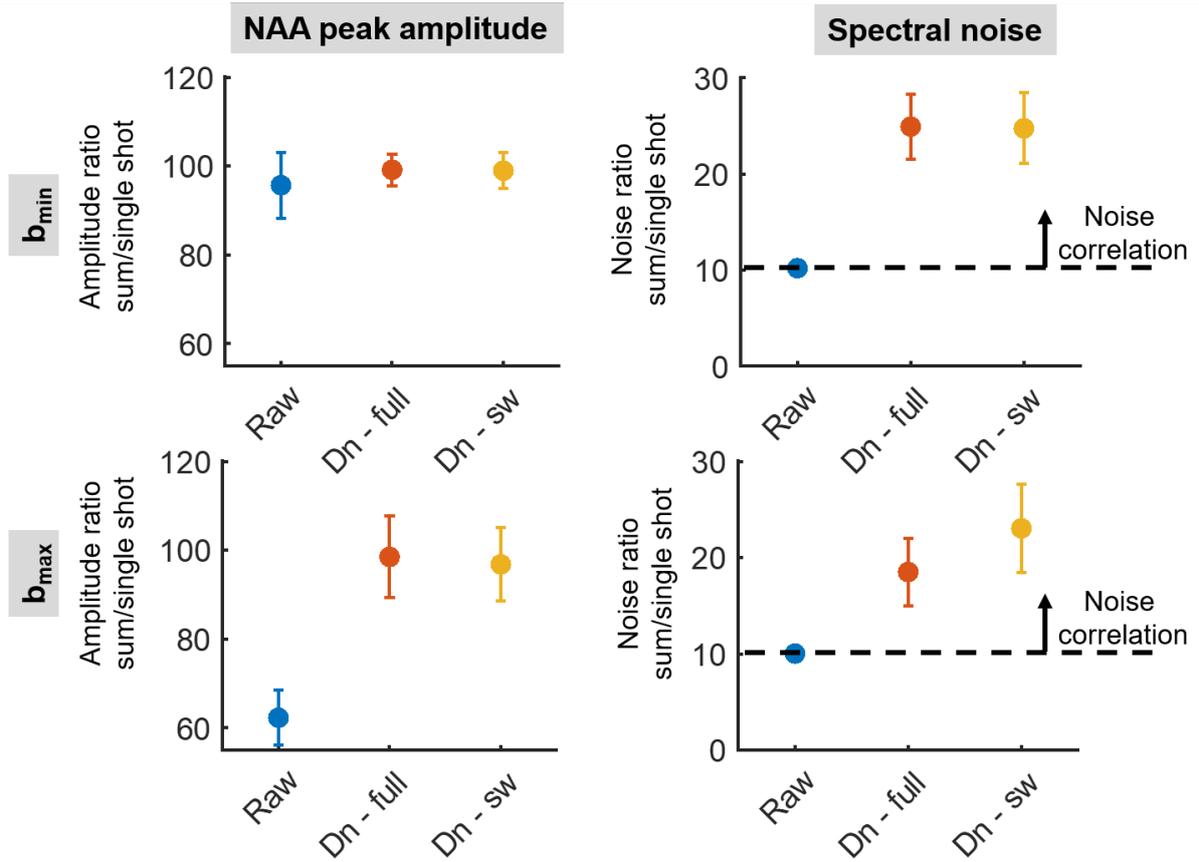

**Figure 5.** Increased correlation in NAA peak amplitude and noise level after denoising, for low and high diffusion weighting. Mean and SD across MC are displayed. Only the correlation between single and multiple shots within one shell is highlighted in this figure.

We further analysed the correlations of the NAA peak amplitude and of the noise between single shots and the sum of NS=100 introduced by MP-PCA (**Figure 5**). The NAA peak amplitude at $b_{min}$ (top left) scales with the number of shots (NS=100), as expected, both for raw and denoised data. At lower SNR ($b_{max}$, bottom left), the NAA peak amplitude on the raw data does not exactly scale with NS because of possible artefacts in the summation, such as improper frequency/phase drifts correction, leading to partially incoherent summation. For the denoised data, the coherent summation property seems to be restored (amplitude ratio close to 100), which can be due to an improved frequency/phase drifts correction after denoising (**Figure 6**) and/or to the creation of more self-similar spectra after rank truncation. The noise level in the raw data displays a ratio that scales with $\sqrt{NS}$, as expected, both for $b_{max}$ and $b_{min}$. For the denoised data, some correlation in the noise across shots is introduced by both denoising strategies, leading to a noise ratio scaling with a factor greater than $\sqrt{NS}$.

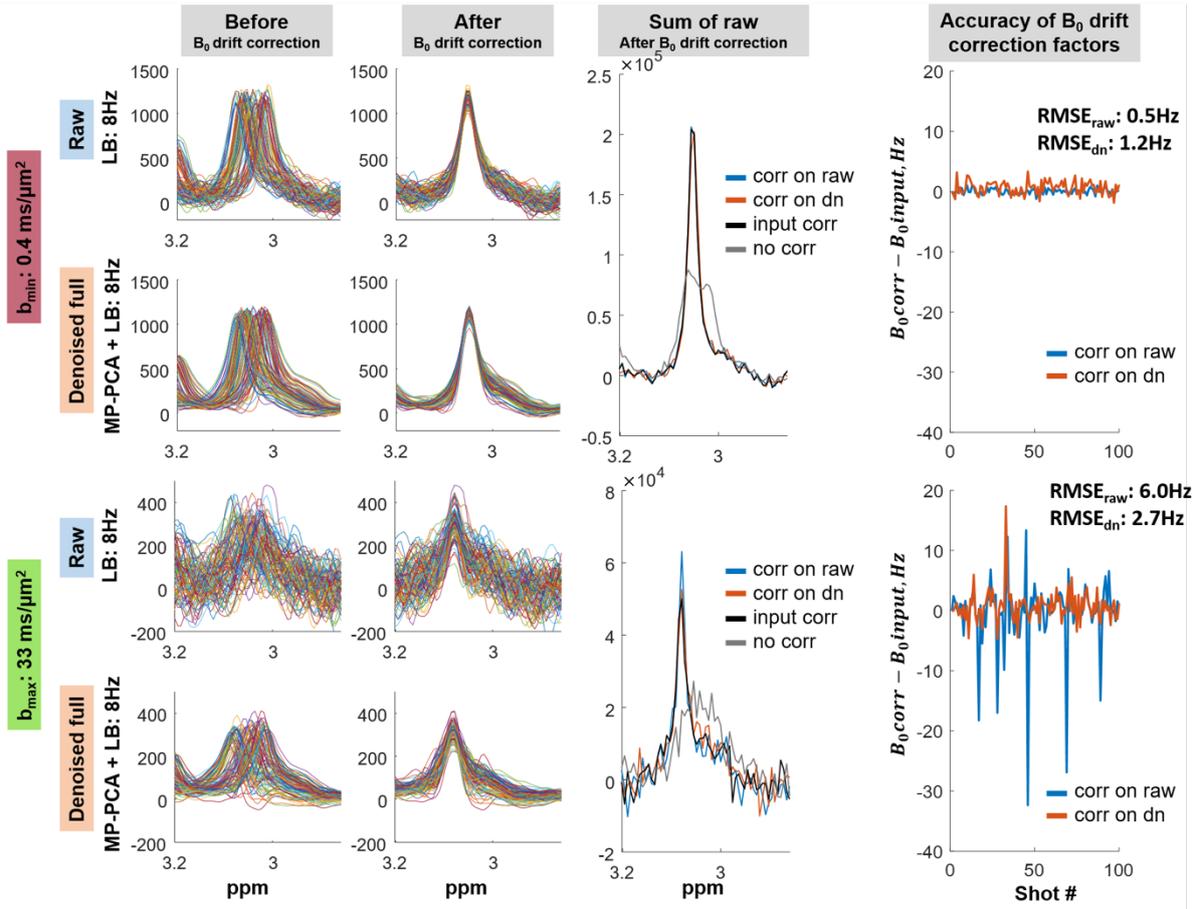

**Figure 6**. Spectral realignment ($B_0$ drift correction) after denoising. The $B_0$ drift correction was performed by aligning the frequency-domain position of the tCr peak to its position on the first spectrum, using a Lorentzian apodization of 8 Hz, on raw and denoised data, for 1 MC iteration (left panels, before/after, for $b_{min}$ and $b_{max}$). The central panel shows the summed raw spectra with corrections derived either from the raw or the MP-PCA data, as compared to the summed raw spectra where the negative input $B_0$ drifts have been applied. Denoising yields no benefit of on $B_0$ drift correction in the case of sufficient SNR (e.g. at $b_{min}$). At low SNR (e.g. $b_{max}$), the summed raw spectra with corrections derived from MP-PCA is closely matching to the one reconstructed from the input $B_0$ drift values, yet with a smaller amplitude than the summed spectra with corrections from the raw data. Denoising before $B_0$ drift correction led to a better accuracy of the B0 drift estimates with respect to the input drifts (right panel) at $b_{max}$, and a worse accuracy at $b_{min}$. $RMSE_{method} = \sqrt{\frac{1}{NS}\sum_{i=1}^{NS}(B_0 corr, method - B_0, input)^2}$.

### 3.1.4. Estimation of metabolite concentrations as a function of b-value

In our post-processing pipeline, denoising was performed before $B_0$ drift correction. This allowed for a more accurate realignment of spectra within each b-value, most noticeably at $b_{max}$ (**Figure 6**): the correction factors derived from the denoised data were closer to ground truth (RMSE: 2.7Hz) compared to the ones derived from the raw data (RMSE: 6.0Hz), although the later yielded a higher amplitude of the summed signal. Metabolites concentrations at $b_{min}$ and $b_{max}$ for all denoising strategies, together with

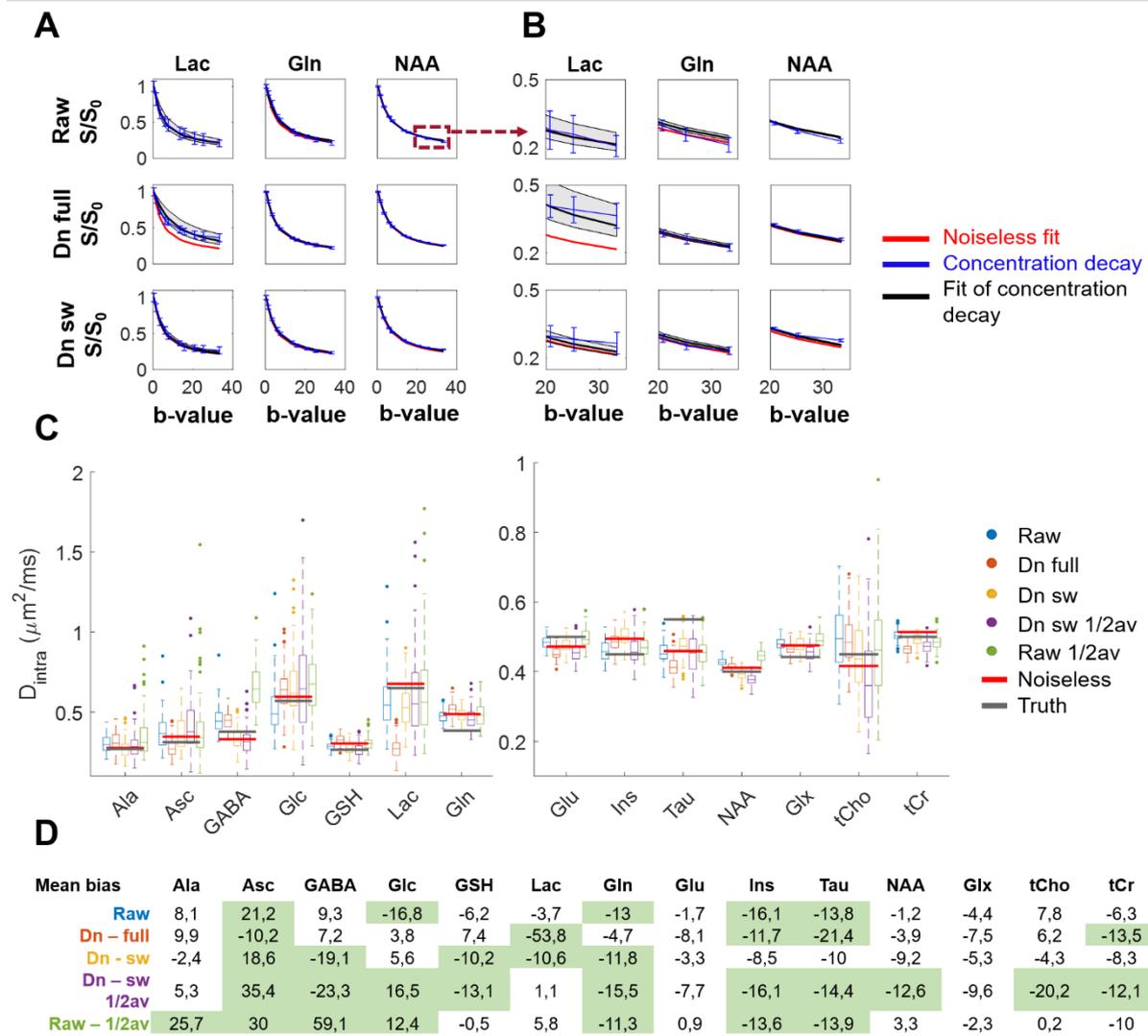

**Figure 7. A:** Representative concentration decay curves for three metabolites: Lac, Gln, NAA, normalized to the concentration at the lowest b-value. Overlaid curves are: mean and SD of concentrations across MC iterations (blue), Callaghan model fit using the mean $D_{intra}$ estimated across MC iterations (black) and Callaghan model fit of the quantified noiseless concentration decay (red). **B:** Zoom-in of panel A for b-values between 20 and 33 ms/μm². **C:** Estimated metabolite $D_{intra}$ from Callaghan's model using raw or denoised data, for various denoising strategies. The values labelled as "truth" represent the diffusion coefficients given as input in the simulations, and the values labelled as "noiseless" represent the LCModel concentrations fit from the noiseless data. **D:** % bias on $D_{intra}$ between all methods and the noiseless fit (($D_{method}$-$D_{noiseless}$)/$D_{noiseless}$). The $D_{intra}$ that differ from the noiseless values by more than ±10% are highlighted in green. Some metabolite-dependant bias on the concentrations and on $D_{intra}$ estimates is either introduced or reduced compared to the raw data after denoising.

Cramer Rao Lower Bounds (CRLB), are presented in **Table S 1**. They highlighted an overall stronger bias introduced by the denoising strategies with respect to the one of the raw data for low-concentrated metabolites, but a weaker one for high-concentrated metabolites, even with strategy 3. Fit precision (CRLB) is strongly improved after denoising for all metabolites. When comparing strategies 1 and 2 on concentration decay curves, their impact was metabolite-dependent (**Figure 7**A-B). In the case of Lac, strategy 1 introduced a systematic bias (overestimated concentration) with respect to noiseless fit, an effect largely mitigated using strategy 2. For Gln, however, both strategies (1 and 2) improved the decay curve accuracy, while no benefit was brought by any of the strategies for NAA.

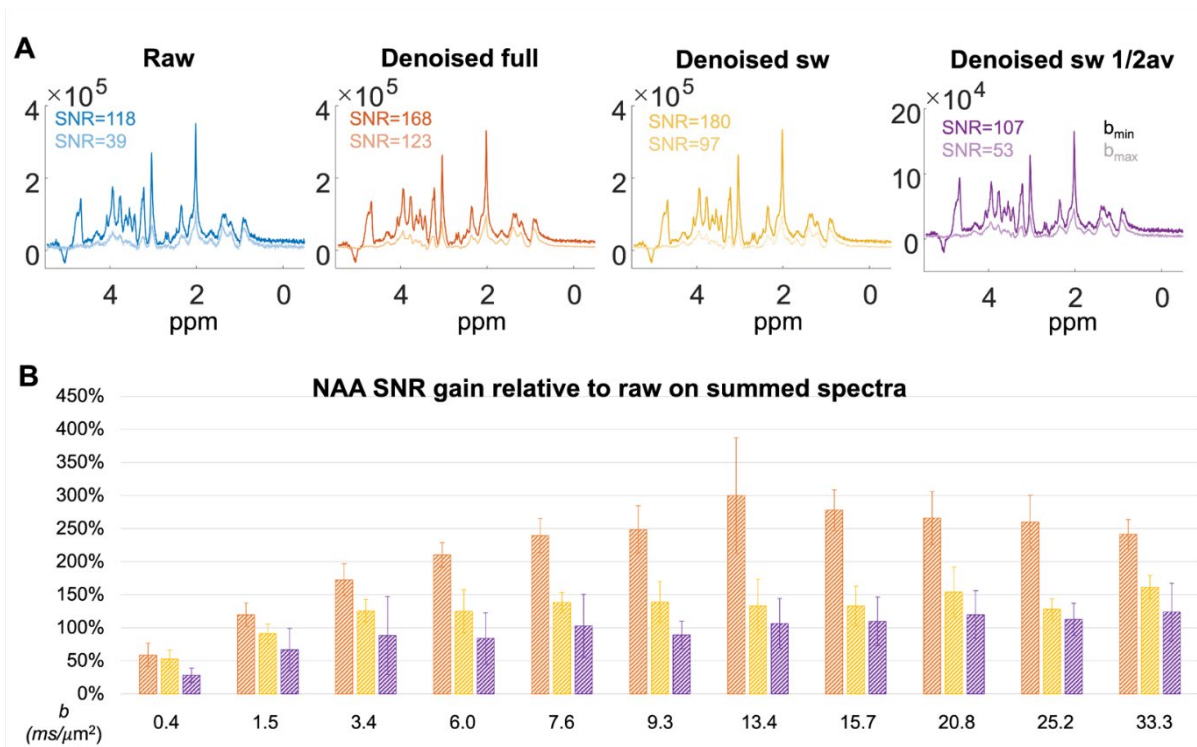

**Figure 8.** Spectral quality and apparent SNR gain, before versus after denoising. **A**: Representative summed spectra for one animal, at low (dark colors) and high (light colors) b-values, based on raw and denoised data, with strategies 1 to 3. NAA singlet SNR is displayed for each case. **B**: Relative apparent SNR gain from the denoising strategies 1 to 3, expressed in % increase compared to the raw data SNR, with mean and SD across animals. Orange: strategy 1, yellow: strategy 2, purple: strategy 3. Increased apparent spectral SNR was observed *in vivo*.

In terms of $D_{intra}$ estimation (**Figure 7**C-D), although p-values highlighted a systematic bias (versus noiseless data, **Table S 3**), strategy 2 led to an improvement in accuracy for some metabolites compared to the raw data and strategy 1 (Ala, tCho, Ins, Tau), a deterioration for some low concentrated metabolites (GABA, GSH, Lac) and similar accuracy for the remaining ones.

The number of outliers was also reduced by all the denoising strategies.

Unfortunately, strategy 3, using half the data (i.e. NS = 50) to assess if the total duration of the scan could be reduced without a significant compromise in accuracy and precision of metabolite concentration and $D_{intra}$, yielded worse or at best similar accuracy and precision of $D_{intra}$ as the full raw data (NS = 100) but also as half the raw data (NS = 50) depending on the metabolite.

### 3.2 *In vivo* data

The same analyses were performed on *in vivo* data from four animals and MP-PCA denoising effects were compared to the ones observed in simulations.

#### 3.2.1 Apparent SNR

The summed spectra for the two extreme b-values before and after denoising using all three strategies are shown in **Figure 8**A. Denoising improved the apparent SNR at all b-values, yet to a smaller extent compared to simulations (**Figure 8**B): on average, the SNR gain is 59% at $b_{min}$ and 241% at $b_{max}$ for strategy 1 and 53% at $b_{min}$ and 161% at $b_{max}$ for strategy 2. The apparent SNR gain follows a similar b-value dependence to the one in simulations, with a maximum for a central b-value for strategy 1 and a constant gain for strategy 2.

3.2.2 Noise properties

For strategies 1 and 2, the noise level on *in vivo* data after denoising was non-uniform across shells, both on the sum and on the single shots (**Figure 9**A), and strategy 2 attenuated this effect: on the summed spectra: $\frac{\sigma_{bmin}}{\sigma_{bmax}} = 2.49$ for strategy 1 and $\frac{\sigma_{bmin}}{\sigma_{bmax}} = 1.87$ for strategy 2. A rank $P = 12$ for strategy 1 and $P \in [4,12]$ for strategy 2 was selected by the MP fit (**Figure 9**B), which was consistent among rats (strategy 1: P=11.5±0.58, strategy 1: $P_{bmin}$=11.25±0.5 and $P_{bmax}$=3.5±1) and similar to the ranks found in simulations (**Figure 3**B).

The spectral residuals for both strategies showed no distinct structure around metabolite frequencies (**Figure 9**C), suggestive of a homogeneous denoising in the spectra.

3.2.3 Estimation of metabolite concentrations as a function of b-value

All denoising strategies yielded similar concentrations and reduced CRLB compared to the raw data for the 6 quantified metabolites at $b_{min}$ and $b_{max}$ (**Table S 2**). Similar trends to those identified in simulations are observed between estimates of $D_{intra}$ from raw and denoised data (**Figure 10**). In the multiple comparison post-hoc test, only tCr $D_{intra}$ showed a significant difference between strategy 1 and 3. For the high-concentrated metabolites (Glu, NAA and tCr), strategy 2 reduced the variability of $D_{intra}$ estimates across animals, as compared to that from the raw data.

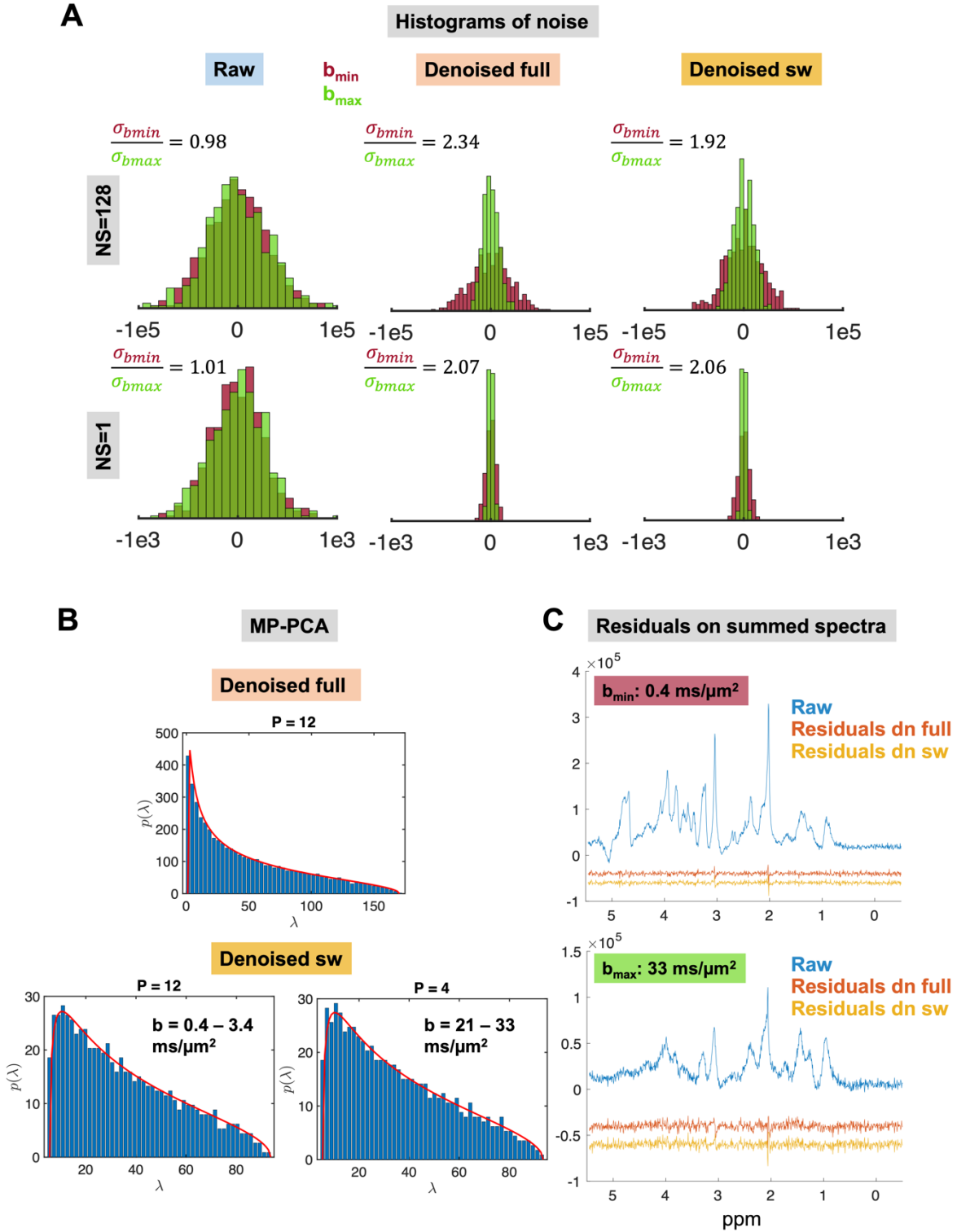

**Figure 9.** MP-PCA denoising performance on the *in vivo* data using strategy 1 and strategy 2. **A:** Histograms of spectral noise for one example animal in the 8.2-10.9 ppm noise-only region, for a single shot (bottom) and for the sum of the 128 shots (top), before and after each denoising strategy, for the smallest (red) and highest (green) b-values. The ratio of the experimental noise level at $b_{min}$ over $b_{max}$ is displayed in each case, averaged over the four animals. Standard deviations across animals associated to the mean ratios displayed: for $NS = 100$, 0.04 (raw), 0.28 (dn full), 0.25 (dn sw), and for $NS = 1$, 0.06 (raw), 0.44 (dn full), 0.65 (dn sw). **B:** MP fit for both strategies. **C:** Residuals between the denoised and raw spectra at the two extreme b-values, after summation of the 128 shots available, shifted downwards for display. The same trends as the ones for simulations can be observed: heterogeneous noise level across shells, increase in apparent SNR with no structure in spectral residuals after denoising, with strategy 2 mitigating some effects of strategy 1.

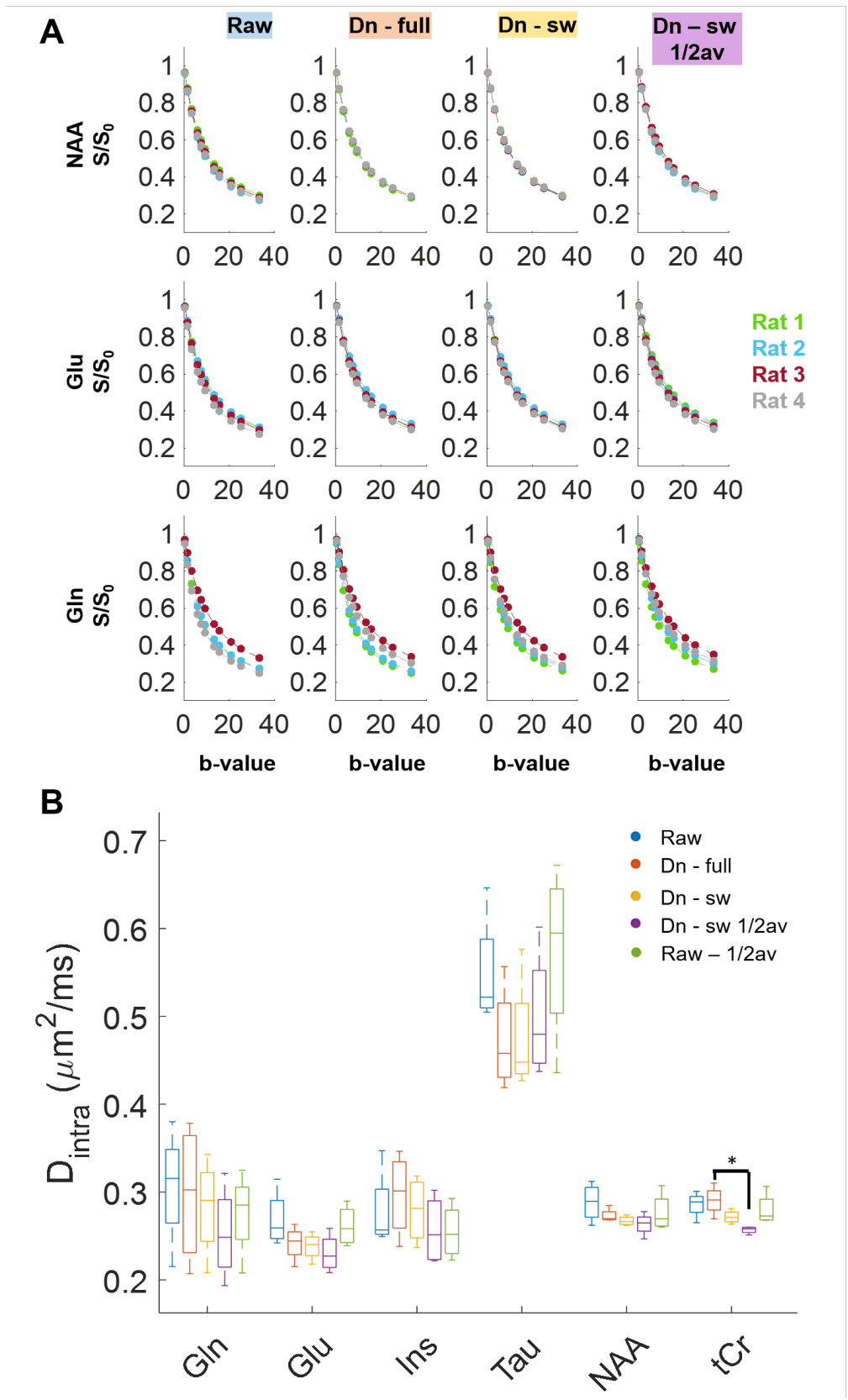

**Figure 10.** Concentration decays after quantification with LCModel, and resulting $D_{intra}$ fit, for raw and denoised data with the three strategies. **A**: Representative decays across b-values for three metabolites: NAA, Glu, Gln, for each animal (circles), with concentrations normalized to the lowest b-value, and individual fits of Callaghan's model (solid line). **B**: Estimated $D_{intra}$ from Callaghan's model for a few metabolites, for all strategies. Raw and denoised data provide similar estimates for most metabolites.

## 4. Discussion

The aim of this work was to evaluate the performance of MP-PCA denoising on synthetic and experimental datasets of single-voxel diffusion-weighted ¹H-MRS comprising spectra at multiple diffusion-weightings (b-values), as compared to conventional averaging across each b-value. We investigated three denoising strategies, comparing their impact on the data structure (apparent SNR increase, spectral residuals, noise correlation), and evaluating their potential for improved diffusion coefficient estimates. Similar characteristics of the denoised spectra were observed between simulations and *in vivo* data (rank P selected by the MP fit, noise ratios between first and last shells, apparent SNR gain, spectral residuals) thus ensuring that conclusions drawn from simulations with respect to the ground truth are relevant for the *in vivo* dataset.

### 4.1 Increased apparent spectral SNR

Simulations revealed that denoising all DW-spectra together significantly improved apparent spectral SNR for each b-value compared to averaging (**Figure 3** and **Figure 4** for simulations and **Figure 8** for *in vivo* data). Remarkably, denoising also provided two valuable features vs averaging.

First, the correction for $B_0$ drifts between individual shots of high b-value shells was more reliable after denoising especially at low SNR (**Figure 6**). Whether a stronger apodization or spectral registration at high b-values could mimic the benefit of denoising prior to $B_0$ drift correction should be further tested with multiple datasets. Interestingly, the correction for phase drifts did not improve after denoising. From this perspective, denoising could be used to determine the optimal frequency drift corrections on individual spectra, and apply it to raw spectra, as previously described in a simpler spectral pattern[44].

Second, the individual spectra after denoising displayed dramatically higher apparent SNR than raw spectra, even at the highest b-value. This single-shot SNR increase, however, results from a correlation with spectra from other shells. Whether this improvement on single shots may benefit other applications where averaging multiple spectra is detrimental, such as functional MRS, where it could provide a boost in temporal resolution, should be the subject of future work.

### 4.2 Strategy 1 versus strategy 2

For large heterogeneities across the dataset to be denoised, such as the extreme case of very low and very high b-values (in our case $b = 0.4$ and up to $b = 33\ ms/\mu m^2$), more leakage from high to low-SNR data is expected after denoising, which may bias high b-values concentration estimates.
Two approaches can however mitigate this effect.
The first approach is to denoise using a sliding-window along b-values, so that the spectra used in each denoising matrix are more similar to each other in terms of SNR. Here we tried a sliding window of three b-values (effectively leading to 3x2xNS rows, accounting for real and imaginary parts of the signal, where NS=100 - 128 are the number of shots acquired for each b-value). While this approach resulted in a more limited noise reduction, especially at high b-values (apparent SNR increase compared to raw of 575% for strategy 1 and 265% for strategy 2 at $b_{max}$ in simulations and 241% for strategy 1 and 161% for strategy 2 at $b_{max}$ *in vivo*), it preserved noise variance better across b-values. The apparent SNR increase is higher in simulations compared to *in vivo* data, possibly owing to sources of non-Gaussian noise or distortions present in raw *in vivo* spectra and absent in simulations. These variations will not be captured in the noise principal components, thus leading to a smaller denoising effect. The number of components retained (P) as signal-carrying decreased across blocks, both in simulations and *in vivo* (**Figure 4**C and **Figure 8**B). One reason is that at high b-values, the variance created by the actual (low SNR) signal is close to the noise floor. An additional hypothesis is that sources of structural/physiological noise in the spectra (e.g. frequency drifts) are more discernible at low b-value than at high b-value and contribute to signal-carrying components.

The second mitigating approach, which remains to be tested, could be to diversify the DW-MRS acquisition scheme not only into multiple b-values, but also directions and diffusion times instead of plain repetitions. While working with only a small range of b-values (with similar SNR) – as for the sliding window above - the denoising matrix construct could nonetheless collect multiple directions and diffusion times. This would also enable to generate large matrices, improving the MP-PCA performance by fulfilling the asymptotic condition of the random matrix theory.

### 4.3 Assessment of denoising quality

One important aspect of MP-PCA denoising is the assumption of Gaussian, constant and uncorrelated entry noise. This assumption can be easily violated for MR imaging in clinical setups where multi-channel receiver coils are recombined using sum-of-squares algorithms, following which the magnitude of the complex signal is retained. In contrast, our *in vivo* preclinical setup was ideal to fulfil this criterion, as the receiver coil was a quadrature circuit whose signals were recombined physically prior to amplification. Each channel (real and imaginary) of the complex signal retained Gaussian noise properties. Despite the apparent SNR increase and homogenous residuals within each shot, some noise correlation across shells introduced by MP-PCA were identified in the current study. Consequently, noise estimation with a prior of uncorrelated Gaussian noise should be avoided in denoised spectra, as well as quality assessment based on noise amplitude, such as CRLB or the fit quality number (FQN)[33]. A bootstrapping approach for the estimation of metabolites concentration uncertainty has been recently proposed[19], where multiple fits of the same spectrum corrupted by correlated noise estimated from the denoised data, are performed.

## 4.4 Noise properties

**Uniform noise level across spectral points:** When comparing single-shell denoising to averaging (**Figure 2**), we observed no residual patterns and the noise was Gaussian across shots and spectral points. Interestingly, in the case of an input matrix with only one signal information, which is not centered (see **Theory** section), and when a rank $P = 1$ is manually selected, a non-uniform variance on spectral points is observed. Concretely, the standard deviation of the spectrum across MC iterations is higher in the metabolite region of the spectrum and smaller in the noise region. This effect was reported in a recent work assessing different low-rank denoising methods for MRSI data[19]. When the matrix is not centered, the only singular value selected will be an estimate of the mean of the input matrix, which might be biased. Remarkably, in the present work, no non-uniform variance across spectral points was observed when denoising a matrix comprised of multi-shell data, even without centering. The high number of principal components selected, $P \sim 11 - 12$, probably mitigates this effect. Similarly, a uniform spectral variance was observed in another FID MRSI study where the MP fit selected $P \sim 20$[21].

**Non-uniform noise level across b-values:** In the present work, the noise level was b-value dependent after MP-PCA denoising (less noise in the high b-value spectra), an effect which was reduced by using a sliding-window across b-values.

**Figure S 2** gives a tentative explanation of the gain in apparent SNR after denoising with strategy 1 reaching a maximum value for intermediate b-values, $b \sim 7.6 \, ms/\mu m^2$ for the simulations (**Figure 4**C) and $b \sim 13.4 \, ms/\mu m^2$ for the *in vivo* data (**Figure 8**B). After MP-PCA, the time evolution of the spectral points in a noise-only region will be reconstructed from one of the first signal-carrying singular vectors in the shot dimension ($U_1$ **Figure S 2**B), representing the overall decay of metabolites across b-values (strongest contribution to the variance). Consequently, the noise points will decay with increasing b-values. Due to the initial positive/negative distribution of these noise points (**Figure S 2**D-E), the noise level (standard deviation of the noise points across a spectral region) will decrease at intermediate b-values and increase again at higher b-values. Meanwhile, the NAA concentration decay is similar for raw and denoised data, which results in a maximum apparent SNR at intermediate b-values. With a similar argument in the other dimension, the first signal-carrying singular vectors in the spectral dimension will represent high SNR spectra ($V_1$ in **Figure S 2**B). The closer the metabolite information to the noise level, the more likely it will be reconstructed from a linear combination of high SNR spectral information, and even more so when $P$ is small. This observation challenges the use of MP-PCA denoising for extracting low-concentrated metabolites information from the noise floor using the entire range of b-values. The sliding window approach can however mitigate these effects, as shown throughout the present work.

The number of principal components retained with strategy 1 was $\sim 11 - 12$ for simulations and *in vivo* data, which was also the rank found when using optimal shrinkage of the principal components[45]. The high number of components was mostly due to the $B_0$ drift distortions which were not corrected for prior to denoising, to the random water residual, and to possible sources of non-Gaussian noise in *in vivo* data. Structural noise, retained as signal component, which has a larger impact on low-b spectra (in particular the water residual which is completely suppressed at high b-values) may therefore serendipitously limit the impact of noise reduction across shells.

### 4.5 Estimation of diffusion coefficients

From the perspective of metabolite quantification, MP-PCA denoising reduced the concentrations CRLB (**Table S 1** and **Table S 2**), representing the lower bounds of the fitting error and assuming Gaussian uncorrelated noise, a prior which is violated after denoising. Simulations showed that denoising based on the full range of b-values could also introduce bias for some metabolite concentration decays, such as lactate (**Figure 7**A-B), and an over-estimation of the concentrations at high b-values compared to the same concentrations on the raw data (**Table S 1**). Interestingly, this over-estimation is not systematic anymore when comparing denoised vs noiseless data: although beyond the scope of this work, this observation highlights some systematic underestimation of concentrations with LCModel for raw data with realistic SNR. The sliding-window approach (strategy 2) introduced less bias on metabolite concentrations at high b-values than the full-range denoising and the raw data for high-concentrated metabolites, in addition to better preserving the noise structure. The observations made on the accuracy and precision of metabolite quantification did not merely translate to the estimations of the free diffusion coefficients $D_{intra}$. Overall, the sliding window-denoising followed the raw data estimates for most metabolites: whether or not bias (>10%) existed in the raw data estimates, the same was observed for strategy 2. The only exceptions are GABA, GSH, Lac, for which more bias was introduced with strategy 2 and Glc, Ins, Tau for which less bias was introduced with strategy 2. However, this performance may depend on the underlying diffusivity values chosen in our simulations. In simulations, the variability across MC iterations was also reduced after denoising (when compared to the raw data with the same number of shots) for all metabolites (**Figure 7**C). Remarkably, *in vivo*, MP-PCA denoising also contributed to reducing the variability in metabolite concentration decay curves across the different rats (which were all part of a homogeneous control group) for some metabolites (NAA, tCr, Glu on **Figure 10**). *In vivo*, the estimated metabolite diffusivities were systematically lower with MP-PCA denoising vs raw data, though the ground truth is not known in this case. This could reflect the systematic under-estimation of the raw data concentrations found in simulations (mentioned above), yielding lower concentration values at the tail of the curve and thus a higher estimated diffusivity. It should also be noted that Callaghan's model of randomly-oriented sticks may not be well-suited to describe the diffusion of certain metabolites *in vivo*, e.g. if they are also extracellular and/or if the radius of the dendrites cannot be assumed to be effectively zero.

In agreement with previous studies, we suggest PCA-denoising for diffusion MRS should be used with caution and we recommend that all effects should be tested in simulations prior to drawing conclusions on *in vivo* data.

## Conclusion

Overall, MP-PCA denoising on DW-MRS data significantly improves apparent SNR but affects noise properties as observed in the present study involving simulations and *in vivo* data. Our results suggest that MP-PCA denoising can be useful to reduce outliers and thus the heterogeneity within an experimental group with little penalty to the diffusivity estimates, to improve the accuracy of the $B_0$ drift correction at high b-value and to improve spectral appearance (apparent SNR gain). Correlations were introduced as a consequence of the rank truncation but uniform variance along the spectrum was preserved due to the matrix centering and the selection of a high rank P by the MP fit (with uncorrected $B_0$ drift prior to denoising). Yet, non-uniform noise levels across b-values were observed (noise "decay"), an effect which was mitigated by using a sliding-window (i.e. denoising more self-similar matrices). Some pitfalls should be avoided such as the appealing apparent benefits for low-concentrated metabolites after denoising when the input matrix is heterogeneous, or the reduced CRLB as a measure of reduced quantification uncertainty. Whether other characteristics of the denoised data could be relevant remains to be tested for each data type.

## Acknowledgements

This work was supported by the Center for Biomedical Imaging of the UNIL, UNIGE, HUG, CHUV, EPFL, the Leenaards and Jeantet Foundations and the SNSF projects no 310030_173222, 310030_201218, the European Union's Horizon 2020 research and innovation program under the Marie Sklodowska-Curie grant agreement No 813120 (INSPiRE-MED). I.O.J. is supported by the SNSF under an Eccellenza grant PCEFP2_194260.
The authors would like to thank Dr. Jelle Veraart, Dr. William Clarke, Dr. Bernard Lanz, Dr. Julien Valette and Dr. Roland Kreis for many interesting discussions on the topic.

# Supplementary material

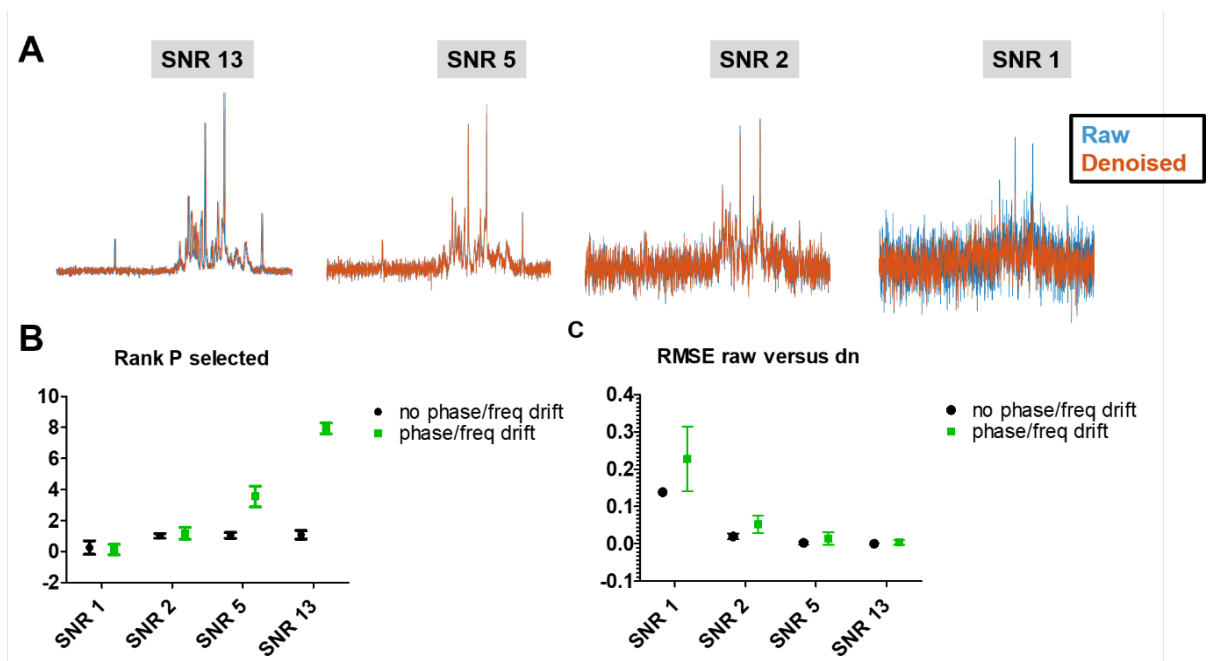

**Figure S 1** - Dependence of denoising performance on the original signal to noise ratio (SNR) of the spectra, and on potential phase and frequency drifts, for a series of 100 individual, non-diffusion-weighted spectra. **A:** Overlap of the raw and denoised spectra (sum over 100 shots) for multiple SNR values, in the absence of phase or frequency drifts. For SNR 1, the spectral information is hardly recovered. From SNR 2 to 13, no or little difference between raw and denoised spectra is observed, even in the presence of phase and frequency drifts, which confirms MP-PCA denoising is similar to averaging in the case of repeated measurements at sufficiently high SNR. **B:** Number of signal-carrying principal components P retained by the MP fit, as a function of SNR and of the presence/absence of phase and frequency drifts. P is expected to be 1 in the general case of repeated measurements only altered by noise between different iterations. In our work, for the single-shell, the concatenation of real and imaginary parts yields a rank 2 before centering, and a rank 1 after centering. In practice, it is on average equal to 1 for SNR 2 to 13 in the absence of drift and phase distortions. When adding drift and phase distortions, P increased with the SNR. The frequency drift plays a larger role than phase in this observation, since it adds structured information in both dimensions of the matrix, which is preserved and rendered as signal-carrying components at high SNR. For SNR 1, P is close to zero and hardly no signal information is retained. **C:** Root mean squared error (RMSE) between the real part of the summed raw versus denoised spectra, where the amplitude of NAA is normalized to one. The RMSE increases with decreased SNR and addition of phase and frequency drifts, as previously observed[19]. For panels B and C, the mean and standard deviations across 100 Monte Carlo iterations are displayed.

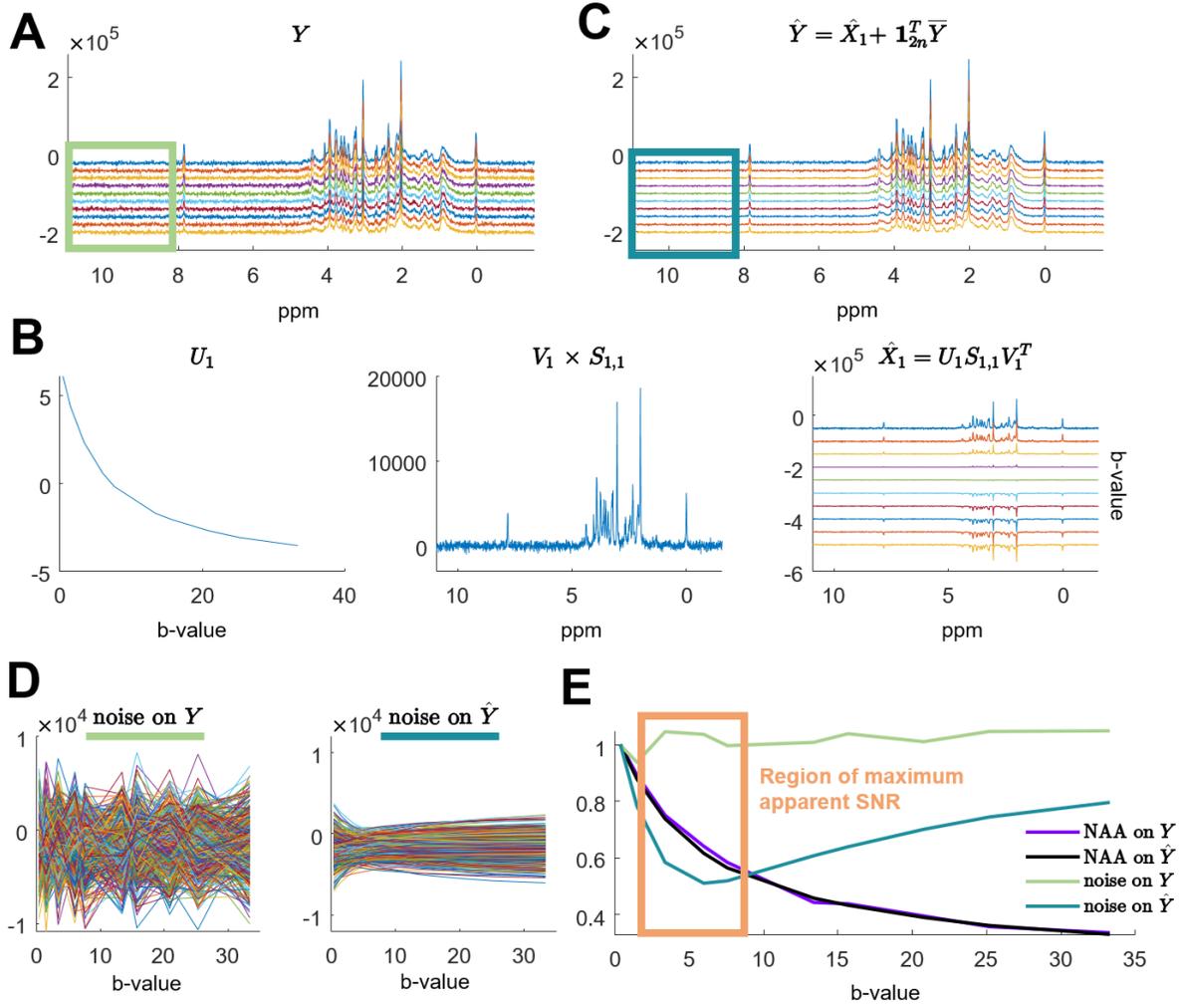

**Figure S 2** - Graphical representation of the effect of MP-PCA on one simulated diffusion dataset denoised with strategy 1. For an easier visualisation of the effects, and for this figure only, MP-PCA has been applied on the real part of spectral matrix (instead of the FID time points, with concatenated real and imaginary parts), and no phase/frequency distortions were introduced in the initial matrix, nor a residual water signal. The matrix is however centered as described above. **A:** Initial matrix Y (1000x2048) where each shell has been summed. **B:** Decomposition of $X = Y - \bar{Y}$ in terms of singular vectors of the shot dimension $U$, and of the spectral dimension $V$. A rank $P = 1$ was selected by the MP fit, such that $\hat{X} = \widehat{X_1} = U_1 S_{1,1} V_1^T$. **C:** Denoised matrix $\hat{Y}$ (1000x2048) where each shell has been summed and the initial mean $\bar{Y}$ reintroduced. **D:** Noise profile across b-values for each spectral point in the noise-only region between 8.2 and 10.9 ppm, shown in panels A and C, before (left) and after (right) denoising. **E:** Standard deviation of noise values from panel D, at each b-value, for raw (light green) and denoised (dark green), together with the decay of the maximum real value of the NAA peak at 2.01ppm. Panel E shows that, in that case, points from a noise-only region, which are reconstructed from a decaying component $U_1$, will also decay. Consequently, the noise level (standard deviation over the points from a noise-only region) after denoising reaches a minimum value at an intermediate b-value (around $b \sim 6 \ ms/\mu m^2$) and grows again at higher b-values due to the initial positive/negative distribution of the noise points. On the other hand, the NAA signal decays similarly before and after denoising, and explains why the apparent SNR reaches a maximum for one of the intermediate b-values, for simulations (**Figure 4**C) and *in vivo* data (**Figure 7**B).

| Simulations | | | | | | | | | | | | | |
|---|---|---|---|---|---|---|---|---|---|---|---|---|---|
| Conc (mmol per kg ww) | | | | | | | CRLB (%) | | | | | | |
| b-value (ms/µm²) | 0,4 | | | | | | b-value (ms/µm²) | 0,4 | | | | | |
| Quantification | Raw | Raw ½ | Dn – strategy 1 | Dn – strategy 2 | Dn – strategy 3 | Noiseless | Quantification | Raw | Raw ½ | Dn – strategy 1 | Dn – strategy 2 | Dn – strategy 3 | Noiseless |
| Ala | 0,675 | 0,663 | 0,669 | 0,683 | 0,696 | 0,669 | Ala | 0,086 | 0,136 | 0,04 | 0,041 | 0,055 | 0,03 |
| Asc | 0,847 | 0,865 | 0,937 | 0,96 | 0,905 | 0,799 | Asc | 0,134 | 0,202 | 0,055 | 0,057 | 0,086 | 0,04 |
| GABA | 1,351 | 1,292 | 1,302 | 1,315 | 1,349 | 1,471 | GABA | 0,058 | 0,089 | 0,03 | 0,03 | 0,04 | 0,02 |
| Glc | 1,424 | 1,61 | 1,376 | 1,359 | 1,346 | 1,529 | Glc | 0,103 | 0,12 | 0,055 | 0,059 | 0,08 | 0,03 |
| GSH | 1,349 | 1,358 | 1,364 | 1,36 | 1,35 | 1,384 | GSH | 0,04 | 0,06 | 0,02 | 0,02 | 0,03 | 0,01 |
| Lac | 0,67 | 0,683 | 0,577 | 0,623 | 0,638 | 0,62 | Lac | 0,084 | 0,128 | 0,043 | 0,042 | 0,059 | 0,03 |
| Gln | 2,495 | 2,531 | 2,403 | 2,385 | 2,41 | 2,529 | Gln | 0,033 | 0,05 | 0,02 | 0,02 | 0,023 | 0,01 |
| Glu | 8,942 | 8,876 | 8,856 | 8,837 | 8,679 | 9,486 | Glu | 0,01 | 0,02 | 0,01 | 0,01 | 0,01 | 0 |
| Ins | 5,226 | 5,388 | 5,105 | 5,164 | 5,141 | 5,571 | Ins | 0,02 | 0,021 | 0,01 | 0,01 | 0,015 | 0,01 |
| Tau | 3,726 | 3,787 | 3,552 | 3,637 | 3,667 | 4 | Tau | 0,03 | 0,036 | 0,02 | 0,02 | 0,02 | 0,01 |
| NAA | 7,704 | 7,603 | 7,651 | 7,643 | 7,586 | 8,057 | NAA | 0,01 | 0,02 | 0,01 | 0,01 | 0,01 | 0 |
| Glu+Gln | 11,437 | 11,407 | 11,26 | 11,223 | 11,089 | 12,014 | Glu+Gln | 0,01 | 0,02 | 0,01 | 0,01 | 0,01 | 0 |
| GPC+PCho | 0,891 | 0,803 | 0,868 | 0,872 | 0,854 | 0,893 | GPC+PCho | 0,073 | 0,105 | 0,034 | 0,036 | 0,05 | 0,02 |
| Cr+PCr | 7,142 | 7,138 | 7,093 | 7,108 | 7,002 | 7,5 | Cr+PCr | 0,01 | 0,013 | 0,01 | 0,01 | 0,01 | 0 |
| Conc (mmol per kg ww) | | | | | | | CRLB (%) | | | | | | |
| b-value (ms/µm²) | 33 | | | | | | b-value (ms/µm²) | 33 | | | | | |
| Quantification | Raw | Raw ½ | Dn – strategy 1 | Dn – strategy 2 | Dn – strategy 3 | Noiseless | Quantification | Raw | Raw ½ | Dn – strategy 1 | Dn – strategy 2 | Dn – strategy 3 | Noiseless |
| Ala | 0,211 | 0,181 | 0,213 | 0,239 | 0,24 | 0,204 | Ala | 0,278 | 0,8 | 0,056 | 0,072 | 0,112 | 0,05 |
| Asc | 0,208 | 0,247 | 0,339 | 0,257 | 0,242 | 0,206 | Asc | 0,724 | 1,625 | 0,067 | 0,238 | 0,452 | 0,1 |
| GABA | 0,247 | 0,15 | 0,33 | 0,394 | 0,4 | 0,423 | GABA | 0,257 | 0,68 | 0,048 | 0,059 | 0,086 | 0,03 |
| Glc | 0,301 | 0,251 | 0,249 | 0,266 | 0,26 | 0,313 | Glc | 0,304 | 0,416 | 0,133 | 0,195 | 0,581 | 0,08 |
| GSH | 0,364 | 0,355 | 0,417 | 0,438 | 0,447 | 0,397 | GSH | 0,128 | 0,195 | 0,03 | 0,039 | 0,054 | 0,03 |
| Lac | 0,163 | 0,171 | 0,2 | 0,16 | 0,158 | 0,131 | Lac | 0,385 | 0,776 | 0,057 | 0,106 | 0,193 | 0,08 |
| Gln | 0,533 | 0,566 | 0,538 | 0,56 | 0,569 | 0,534 | Gln | 0,144 | 0,212 | 0,03 | 0,042 | 0,062 | 0,03 |
| Glu | 2,089 | 2,007 | 2,163 | 2,259 | 2,252 | 2,3 | Glu | 0,042 | 0,07 | 0,01 | 0,019 | 0,02 | 0,01 |
| Ins | 1,176 | 1,194 | 1,201 | 1,235 | 1,255 | 1,274 | Ins | 0,058 | 0,087 | 0,02 | 0,028 | 0,034 | 0,01 |
| Tau | 0,919 | 0,914 | 0,926 | 0,945 | 0,961 | 1,064 | Tau | 0,08 | 0,124 | 0,022 | 0,03 | 0,043 | 0,02 |
| NAA | 1,784 | 1,671 | 1,994 | 2,11 | 2,074 | 2,043 | NAA | 0,033 | 0,053 | 0,01 | 0,017 | 0,02 | 0,01 |
| Glu+Gln | 2,622 | 2,574 | 2,701 | 2,819 | 2,822 | 2,843 | Glu+Gln | 0,041 | 0,064 | 0,01 | 0,016 | 0,02 | 0,01 |
| GPC+PCho | 0,203 | 0,188 | 0,201 | 0,229 | 0,247 | 0,24 | GPC+PCho | 0,275 | 0,375 | 0,064 | 0,089 | 0,127 | 0,05 |
| Cr+PCr | 1,691 | 1,722 | 1,712 | 1,765 | 1,749 | 1,743 | Cr+PCr | 0,031 | 0,048 | 0,01 | 0,011 | 0,02 | 0,01 |

**Table S 1 -** Concentrations and Cramer-Rao Lower Bounds (CRLB) for simulations (mean over MC iterations) at the lowest and highest b-values, for every method and every reliably quantified metabolite. The concentrations that differ from the noiseless values by more than ±10% are highlighted in green (bias: (Conc$_{method}$-Conc$_{noiseless}$)/Conc$_{noiseless}$). In the CRLB tables, the values above 20% are highlighted in blue (fit quality). Denoising expectedly reduces the CRLBs but does not overall improve the quantification bias compared to raw data.

| In vivo | | | | | | | | | | | |
|---|---|---|---|---|---|---|---|---|---|---|---|
| **Conc (mmol per kg ww)** | | | | | | **CRLB (%)** | | | | | |
| b-value (ms/µm²) | 0,4 | | | | | b-value (ms/µm²) | 0,4 | | | | |
| Quantification | Raw | Raw ½ | Dn – strategy 1 | Dn – strategy 2 | Dn – strategy 3 | Quantification | Raw | Raw ½ | Dn – strategy 1 | Dn – strategy 2 | Dn – strategy 3 |
| Gln | 5,098 | 4,791 | 5,027 | 5,038 | 4,857 | Gln | 0,045 | 0,058 | 0,035 | 0,035 | 0,048 |
| Glu | 14,895 | 14,443 | 14,83 | 14,906 | 14,495 | Glu | 0,02 | 0,025 | 0,02 | 0,02 | 0,023 |
| Ins | 9,134 | 8,911 | 9,259 | 9,233 | 8,953 | Ins | 0,025 | 0,033 | 0,02 | 0,02 | 0,028 |
| Tau | 9,507 | 9,631 | 9,182 | 9,238 | 9,507 | Tau | 0,028 | 0,033 | 0,02 | 0,02 | 0,028 |
| NAA | 7,434 | 7,288 | 7,369 | 7,406 | 7,299 | NAA | 0,013 | 0,018 | 0,01 | 0,01 | 0,015 |
| Cr+PCr | 13,424 | 13,163 | 13,547 | 13,474 | 13,18 | Cr+PCr | 0,01 | 0,015 | 0,01 | 0,01 | 0,013 |
| **Conc (mmol per kg ww)** | | | | | | **CRLB (%)** | | | | | |
| b-value (ms/µm²) | 33 | | | | | b-value (ms/µm²) | 33 | | | | |
| Quantification | Raw | Raw ½ | Dn – strategy 1 | Dn – strategy 2 | Dn – strategy 3 | Quantification | Raw | Raw ½ | Dn – strategy 1 | Dn – strategy 2 | Dn – strategy 3 |
| Gln | 1,025 | 1,082 | 1,293 | 1,393 | 1,399 | Gln | 0,2 | 0,23 | 0,065 | 0,068 | 0,08 |
| Glu | 4,39 | 4,528 | 4,905 | 5,44 | 5,049 | Glu | 0,053 | 0,065 | 0,023 | 0,023 | 0,03 |
| Ins | 1,936 | 2,22 | 2,462 | 2,64 | 2,581 | Ins | 0,095 | 0,1 | 0,035 | 0,04 | 0,048 |
| Tau | 1,17 | 1,141 | 1,554 | 1,771 | 1,661 | Tau | 0,188 | 0,203 | 0,065 | 0,06 | 0,078 |
| NAA | 2,105 | 2,293 | 2,201 | 2,511 | 2,341 | NAA | 0,04 | 0,045 | 0,023 | 0,023 | 0,025 |
| Cr+PCr | 3,488 | 3,739 | 3,934 | 4,356 | 4,132 | Cr+PCr | 0,04 | 0,048 | 0,015 | 0,02 | 0,023 |

**Table S 2 -** Concentrations and Cramer-Rao Lower Bounds (CRLB) for *in vivo* data (mean over animals, referenced to water at $b = 0.4\ ms/µm^2$) at the lowest and highest b-values, for every method and every reliably quantified metabolite. In the CRLB tables, the values above 20% are highlighted in blue (fit quality). Denoising expectedly reduces the CRLBs. Metabolite concentrations for *in vivo* data at all b-values are referenced to the water signal at $b = 0.4\ ms/µm^2$. They are therefore overestimated even at low b-values because of a faster water diffusivity compared to metabolites. This referencing neither affects the decay of metabolites relative to the first value, nor the difference between the different metabolites' decay because the same water file was used for every metabolite and every b-value.

|  | Ala | Asc | GABA | Glc | GSH | Lac | Gln | Glu | Ins | Tau | NAA | Glu+Gln | GPC+PCho | Cr+PCr | P-value |
|---|---|---|---|---|---|---|---|---|---|---|---|---|---|---|---|
| Raw | | | | | | | | | | | | | | | NS: >0,05 |
| Dn – full | | | | | | | | | | | | | | | *: <0,05 |
| Dn – sw | | | | | | | | | | | | | | | **: <0,01 |
| Dn – sw 1/2av | | | | | | | | | | | | | | | ***: <0,001 |
| Raw – 1/2av | | | | | | | | | | | | | | | ****: <0,0001 |

**Table S 3 -** Statistical differences between $D_{intra}$ in the MC study with respect to the fit value on noiseless data, reported as p-values from 1-way ANOVA with Dunnett's post-hoc test. The statistical significance for the high-concentrated metabolites is misleading. It results from low CRLB for these (well-quantified) metabolites together with a systematic bias introduced by LCModel.